\documentclass[12pt]{iopart}
\usepackage{dcolumn,graphicx,color,booktabs,microtype,afterpage}
\usepackage{float}
\usepackage{amssymb}
\usepackage{url}  
\usepackage[charter,greekuppercase=italicized]{mathdesign}
\usepackage{sidecap}
\usepackage[mathlines]{lineno}

\usepackage{color}
\usepackage[colorlinks,plainpages=false,linkcolor=blue,urlcolor=blue,citecolor=blue,pdfpagemode=UseNone,pdfstartview=FitBH]{hyperref}
\graphicspath{{./}{figure/}}

\begin{document}
	
	\makeatletter\renewcommand{\ps@plain}{%
		\def\@evenhead{\hfill\itshape\rightmark}%
		\def\@oddhead{\itshape\leftmark\hfill}%
		\renewcommand{\@evenfoot}{\hfill\small{--~\thepage~--}\hfill}%
		\renewcommand{\@oddfoot}{\hfill\small{--~\thepage~--}\hfill}%
	}\makeatother\pagestyle{plain}

\title[Giant MR and THE in the EuGa$_4$ antiferromagnet]{Giant magnetoresistance and topological Hall effect in the EuGa$_4$ antiferromagnet} 
\author{H\ Zhang$^{1}\dagger$, X\ Y\ Zhu$^{1}\dagger$, Y\ Xu$^{1,*}$, D\ J\ Gawryluk$^2$, W\ Xie$^3$, S L Ju$^4$, M\ Shi$^4$, T~Shiroka$^{5,6}$, Q\ F\ Zhan$^7$, E Pomjakushina$^{3,*}$, and T\ Shang$^{7,*}$}
\address{$^1$School of Physics and Electronic Science, East China Normal University, Shanghai 200241, China}
\address{$^2$Laboratory for Multiscale Materials Experiments, Paul Scherrer Institut, CH-5232 Villigen PSI, Switzerland}
\address{$^3$DESY, Notkestra$\beta$e 85, D-22607 Hamburg, Germany}
\address{$^4$Swiss Light Source, Paul Scherrer Institut, CH-5232 Villigen PSI, Switzerland}
\address{$^5$Laboratory for Muon-Spin Spectroscopy, Paul Scherrer Institut, CH-5232 Villigen PSI, Switzerland}
\address{$^6$Laboratorium f\"ur Festk\"orperphysik, ETH Z\"urich, CH-8093 Zurich, Switzerland}
\address{$^7$Key Laboratory of Polar Materials and Devices (MOE), School of Physics and Electronic Science, East China Normal University, Shanghai 200241, China}

\eads{\mailto{yxu@phy.ecnu.edu.cn}, \mailto{ekaterina.pomjakushina@psi.ch}, \mailto{tshang@phy.ecnu.edu.cn}}	
	
\date{\today}
	
\begin{abstract}
We report on systematic temperature- and magnetic field-dependent studies of the EuGa$_4$ binary compound, 
which crystallizes in a centrosymmetric tetragonal BaAl$_4$-type 
structure with space group $I4/mmm$. The electronic properties of EuGa$_4$ single crystals, with an antiferromagnetic (AFM) transition at $T_\mathrm{N} \sim 16.4$\,K, 
were characterized via electrical resistivity and magnetization measurements. 
A giant nonsaturating magnetoresistance was observed at low temperatures, 
reaching $\sim 7 \times 10^4$\,\% at 2\,K in a magnetic field of 9\,T. 
In the AFM state, EuGa$_4$ 
undergoes a series of me\-ta\-mag\-net\-ic transitions in an applied magnetic field, 
clearly manifested 
in its field-dependent electrical resistivity. 
Below $T_\mathrm{N}$, in the $\sim$4--7\,T field range, we observe also a clear hump-like anomaly in the Hall resistivity which is part of the anomalous Hall resistivity.
We attribute such a hump-like feature to the topological Hall effect,  
usually occurring in noncentrosymmetric materials known to host 
topological spin textures (as e.g., magnetic skyrmions). 
Therefore, the family of materials with a tetragonal BaAl$_4$-type structure, to which EuGa$_4$ and EuAl$_4$ belong, seems to comprise suitable candidates on which one can study the interplay among correlated-electron phenomena (such as charge-density wave or exotic magnetism) with topological spin textures and topologically nontrivial bands.
\end{abstract}
		
%
%

%
\submitto{\JPCM}

\footnotetext{These authors contributed equally}
\maketitle
%
%

	
\section{Introduction}
Recently, the study of unconventional Hall effect, including 
its anomalous-, topological-, and spin-Hall variants, has become one of the preferred
methods for investigating the interplay between magnetism and topology 
in quantum materials~\cite{Nagaosa2010,Sinova2015,neubauer_topological_2009}.
Generally, the Hall effect involves the deflection of 
charge-carrier trajectories by the Lorentz force. In addition to an 
external magnetic field (the origin of the classical Hall effect), 
other sources of the Lorentz force can also be the effective 
fields associated with a nonzero Berry curvature. To better illustrate this, in general,
the Hall resistivity $\rho_{xy}$ can be written as  
$\rho_{xy} = \rho_{xy}^O + \rho_{xy}^A$, where $\rho_{xy}^O$ and $\rho_{xy}^A$ represent the ordinary- and the anomalous Hall resistivity, respectively. The anomalous Hall effect 
is usually observed in magnetic materials with a finite magnetization 
(e.g., ferromagnets or ferrimagnets), which is due to an intrinsic Karplus-Luttinger mechanism, and extrinsic mechanisms like skew scattering and side jump~\cite{Nagaosa2010}.
A finite $\rho_{xy}^A$ due to to\-po\-lo\-gi\-cal\-ly nontrivial 
momentum-space features, e.g., Dirac- or Weyl points, currently 
attracting an intense research interest~\cite{onoda_anomalous_2004,nakatsuji_large_2015,liang_ultrahigh_2015,liang_anomalous_2017,liang_anomalous_2018,suzuki_large_2016,xu_unconventional}. 
For instance, Mn$_3$Sn, Mn$_3$Ge, and Mn$_3$Ir noncollinear antiferromagnets~\cite{nakatsuji_large_2015,ikhlas_large_2017,chen_anomalous_2014,nayak_large_2016}, 
nonmagnetic- TaAs, TaP, and NbP~\cite{caglieris_anomalous_2018,watzman_dirac_2018}  and magnetic- GdPtBi and YbPtBi~\cite{suzuki_large_2016,guo_evidence_2018} Weyl semimetals, 
Cd$_3$As$_2$ and ZrTe$_5$ Dirac semimetals~\cite{liang_ultrahigh_2015,liang_anomalous_2017,liang_anomalous_2018}, all exhibit $\rho_{xy}^A$ in wide temperature- and magnetic field ranges.

The $\rho_{xy}^A$ term can be further split into two sub-types
(owing to their different origins), i.e., $\rho_{xy} = \rho_{xy}^O + \rho_{xy}^{A'} + \rho_{xy}^T$. 
Here, $\rho_{xy}^{A'}$ represents the conventional anomalous Hall term, 
mostly determined by the electrical resistivity and magnetization.
The second term $\rho_{xy}^T$ indicates a topological Hall term. The topological Hall effect (THE) is considered to be the hallmark of spin textures 
with a finite scalar spin chirality in real space.
Such topological spin textures usually exhibit a nonzero Berry phase, 
here acting as an effective magnetic field, which gives rise 
to the topological Hall resistivity $\rho_{xy}^T$. The THE has been 
frequently observed in magnetic materials with non-coplanar spin textures, such as skyrmions~\cite{neubauer_topological_2009,gayles_dzyaloshinskii-moriya_2015,lee_unusual_2009,kanazawa_large_2011,li_robust_2013,franz_real-space_2014,huang_extended_2012,schulz_emergent_2012,qin_emergence_2019,matsuno_interface-driven_2016,kurumaji_skyrmion_2019}, hedgehogs~\cite{kanazawa_critical_2016,fujishiro_topological_2019}, hopfions~\cite{gobel_topological_2020}, merons~\cite{puphal_topological_2020}, and magnetic bubbles~\cite{vistoli_giant_2019}. 
Among the notable examples in this regard are the noncentrosymmetric MnSi and analogous compounds~\cite{neubauer_topological_2009,gayles_dzyaloshinskii-moriya_2015,kanazawa_large_2011,franz_real-space_2014}, where $\rho_{xy}^T$ is caused by magnetic skyrmions.

To date, the topological Hall effect has been studied mostly in 
transition-metal compounds with a noncentrosymmetric crystal structure~\cite{neubauer_topological_2009,gayles_dzyaloshinskii-moriya_2015,kanazawa_large_2011,franz_real-space_2014}.
Recently, such studies have been extended to rare-earth magnetic 
compounds with a centrosymmetric crystal structure~\cite{kurumaji_skyrmion_2019,Hirschberger2019,Khanh2020}. 
Yet, magnetic materials with a centrosymmetric crystal 
structure that still can host magnetic skyrmions are rare. 
In noncentrosymmetric materials, skyrmions can be stabilized by the Dzyaloshinskii-Moriya interaction~\cite{muhlbauer_skyrmion_2009,yu_near_2011,yu_real-space_2010,seki_observation_2012,kezsmarki_ne-type_2015,tokunaga_new_2015,Seki2012}. Since this is absent in centrosymmetric materials, 
different mechanisms, including magnetic frustration and fluctuation or 
the competition between magnetic interactions and magnetic 
anisotropies, have been proposed to lead to the formation of skyrmions~\cite{kurumaji_skyrmion_2019,Hirschberger2019,Khanh2020,Ghimire2020,Batista2016,li_large_2019}. 
Nevertheless, such mechanisms cannot account for all the cases
where skyrmions are observed in centrosymmetric materials. 
Hence, their origin is not yet fully understood and requires further investigations.

The discovery of nontrivial band topology and large magnetoresistance (MR) 
in the prototype compound BaAl$_4$ has stimulated 
considerable interest in this family of materials~\cite{wang_crystalline_2021}.
Similar to BaAl$_4$, also BaGa$_4$ exhibits metallic behavior without 
undergoing any phase transitions, while SrAl$_4$ shows a 
charge-density-wave (CDW) and a structural phase transition at 
$T_\mathrm{CDW} \sim 250$\,K and $T_\mathrm{S} \sim 90$\,K, respectively~\cite{Nakamura2016}. 
Both compounds are also expected to exhibit topological features. 
Unlike these nonmagnetic materials, upon replacing Ba (or Sr) 
with Eu, the $4f$ electrons bring new intriguing aspects to the topology, as clearly illustrated by our recent work on EuAl$_4$~\cite{EuAl4_PRB}.
EuAl$_4$ exhibits coexisting antiferromagnetic- (AFM) and CDW orders 
with onset temperatures of  
$T_\mathrm{N} \sim 15.6$\,K and $T_\mathrm{CDW} \sim 140$\,K~\cite{EuAl4_PRB,araki_charge_2013,nakamura_unique_2014,nakamura_transport_2015,shimomura_lattice_2019,Kobata2016} 
and undergoes a series of metamagnetic transitions in the AFM state~\cite{EuAl4_PRB,nakamura_transport_2015}. 
Within the $\sim$1--2.5\,T field range, a clear hump-like anomaly was 
observed in the Hall resistivity, most likely a manifestation of 
THE~\cite{EuAl4_PRB}. Hence, EuAl$_4$ represents a rare case where the 
topological Hall effect not only arises in a centrosymmetric structure, but it also coexists with CDW order.

EuGa$_4$ is also an antiferromagnet (with $T_\mathrm{N} \sim 16.5$\,K), 
whose CDW order starts to emerge under applied 
pressure close to 0.75\,GPa, with $T_\mathrm{CDW}$ reaching $\sim 175$\,K 
at 2.3\,GPa~\cite{nakamura_transport_2015,nakamura_magnetic_2013}. Although 
EuAl$_4$ and EuGa$_4$ share similar magnetic properties, much less is 
known about EuGa$_4$, in particular, with regard to its 
topological transport properties. 
Most of the previous work on EuGa$_4$ has focused on its
temperature-dependent aspects, while its electrical transport 
properties under applied magnetic field have been somewhat 
overlooked~\cite{Kobata2016,nakamura_magnetic_2013}. Here, by 
investigating the temperature- and field-dependent electrical 
resistivity and magnetization of EuGa$_4$, we report 
the observation of (i) a giant nonsaturating MR,
whose origin is most likely attributed to
a nontrivial band structure; 
(ii) a hump-like anomaly in the Hall resistivity, most likely attributed to the 
topological Hall effect induced by topological spin textures.

\section{Experimental details} 	
Single crystals of EuGa$_4$ were grown by a molten Ga flux method. 
High purity Eu rods (Alfa Aesar, 99.9\%) and Ga ingots (Alfa Aesar, 99.999\% ) 
in a ratio of 1:9 were loaded in an alumina crucible and sealed in a quartz ampule.  
Then, the quartz ampule was heated up to 750$^\circ$C at a rate of 
200$^\circ$C/h. After annealing at this temperature for more than
10\,h, the ampule was slowly cooled down to 400$^\circ$C at a rate 
of 1$^\circ$C/h. After removing the excess Ga flux using a centrifuge, centimetre-sized 
single crystals were obtained [see example in the inset of 
figure~\ref{fig:XRD}(b)].

The phase purity and the orientation of EuGa$_4$ crystals were checked by x-ray diffraction (XRD) measurements 
using a Bruker D8 diffractometer. The electrical resistivity and 
magnetization were measured in a Quantum Design MPMS and PPMS on oriented single crystals with typical dimension of 3 $\times$ 1 $\times$ 0.25 mm$^3$, respectively.  
For the magnetization measurements, the magnetic field was applied along the $c$-axis. For the resistivity measurements, the electric current (5\,mA) was applied in the $ab$-plane, while the magnetic field was applied along the $c$-axis. The field-dependent electrical resistivity 
was investigated using both positive and negative magnetic fields. To avoid spurious resistivity contributions due to misaligned Hall probes, the longitudinal contribution to the Hall resistivity $\rho_{xy}$, was removed by an anti-symmetrization procedure, i.e., 
$\rho_{xy}(H) = [\rho_{xy}(H) - \rho_{xy}(-H)]/2$. 
Similarly, in case of longitudinal electrical resistivity $\rho_{xx}$ measurements, the spurious transverse contribution 
was removed by a symmetrization procedure, i.e., 
$\rho_{xx}(H) = [\rho_{xx}(H) + \rho_{xx}(-H)]/2$.

\section{Results and discussion} 
%
\begin{figure*}
	\includegraphics[width=0.9\textwidth]{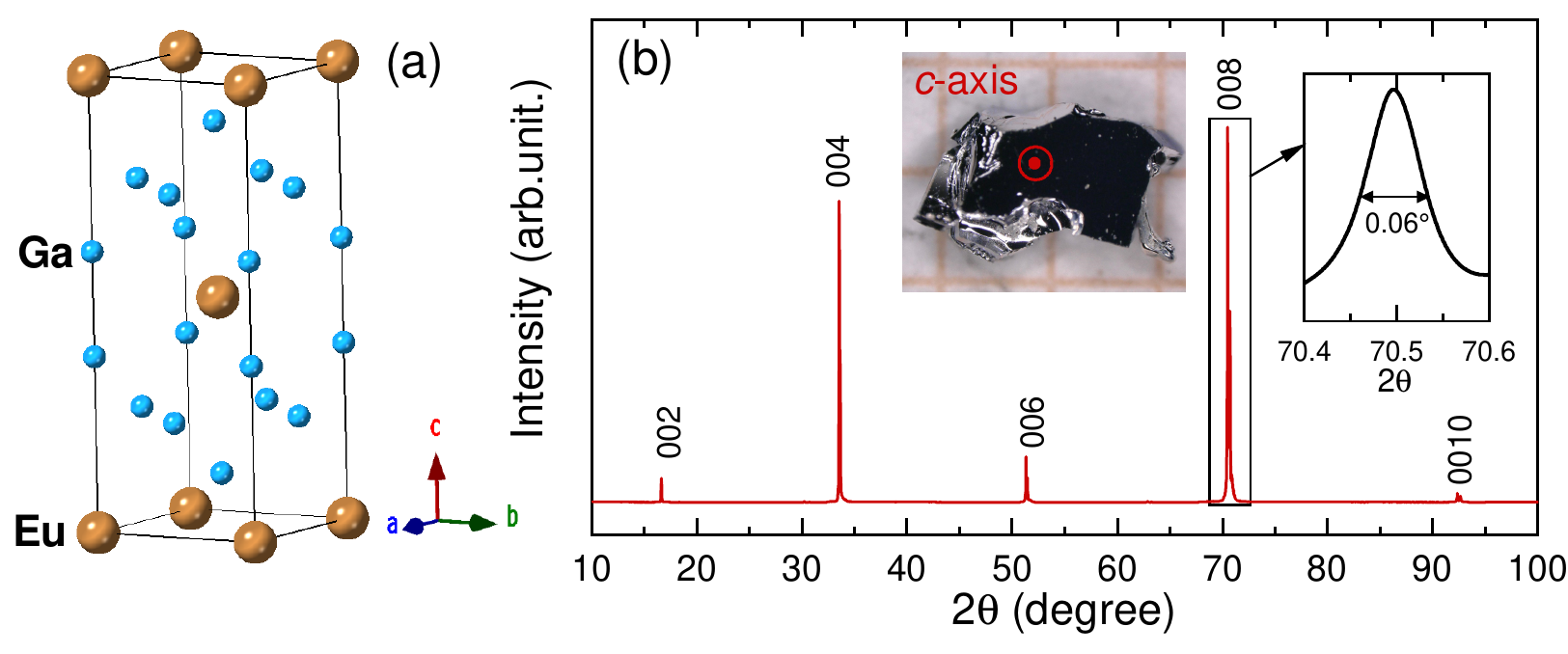}
	\centering
	\vspace{-2ex}%
	\caption{\label{fig:XRD}(a) Crystal structure of EuGa$_4$. (b) XRD pattern of an
	EuGa$_4$ single crystal. The inset in (b) shows a picture 
	of the crystal, whose $c$-axis is normal to the  
	plane. The rather narrow (008) reflection, shown enlarged in the 
	inset, indicates a good sample quality.}
\end{figure*}
%

Figure~\ref{fig:XRD}(a) shows the crystal structure of EuGa$_4$, 
where three Ga- and one Eu-layers stack alternatively along the $c$-axis. 
No foreign phases could be identified in the powder XRD pattern, thus
indicating that the obtained EuGa$_4$ crystals are in a pure phase. 
The XRD pattern is well indexed by a tetragonal crystal structure with space group 
$I4/mmm$ (No.\ 139), fairly typical of 
many binary- and ternary derivative compounds~\cite{Kneidinger2014}, 
as e.g., iron-based superconductors and heavy-fermion compounds.  
The crystal orientation was determined from XRD patterns of EuGa$_4$ 
single crystals. As shown in figure~\ref{fig:XRD}(b), only (00$l$) 
reflections could be detected for the crystal depicted in the inset, thus implying that its
$c$-axis is perpendicular to the EuGa$_4$ crystal plane. The good quality of 
EuGa$_4$ crystals was also confirmed by a relatively small full width 
at half maximum (FWHM), here $\sim 0.06^\circ$ for the (008)-reflection, as shown in the inset of figure~\ref{fig:XRD}(b).

\begin{figure*}
	\includegraphics[width=0.5\textwidth]{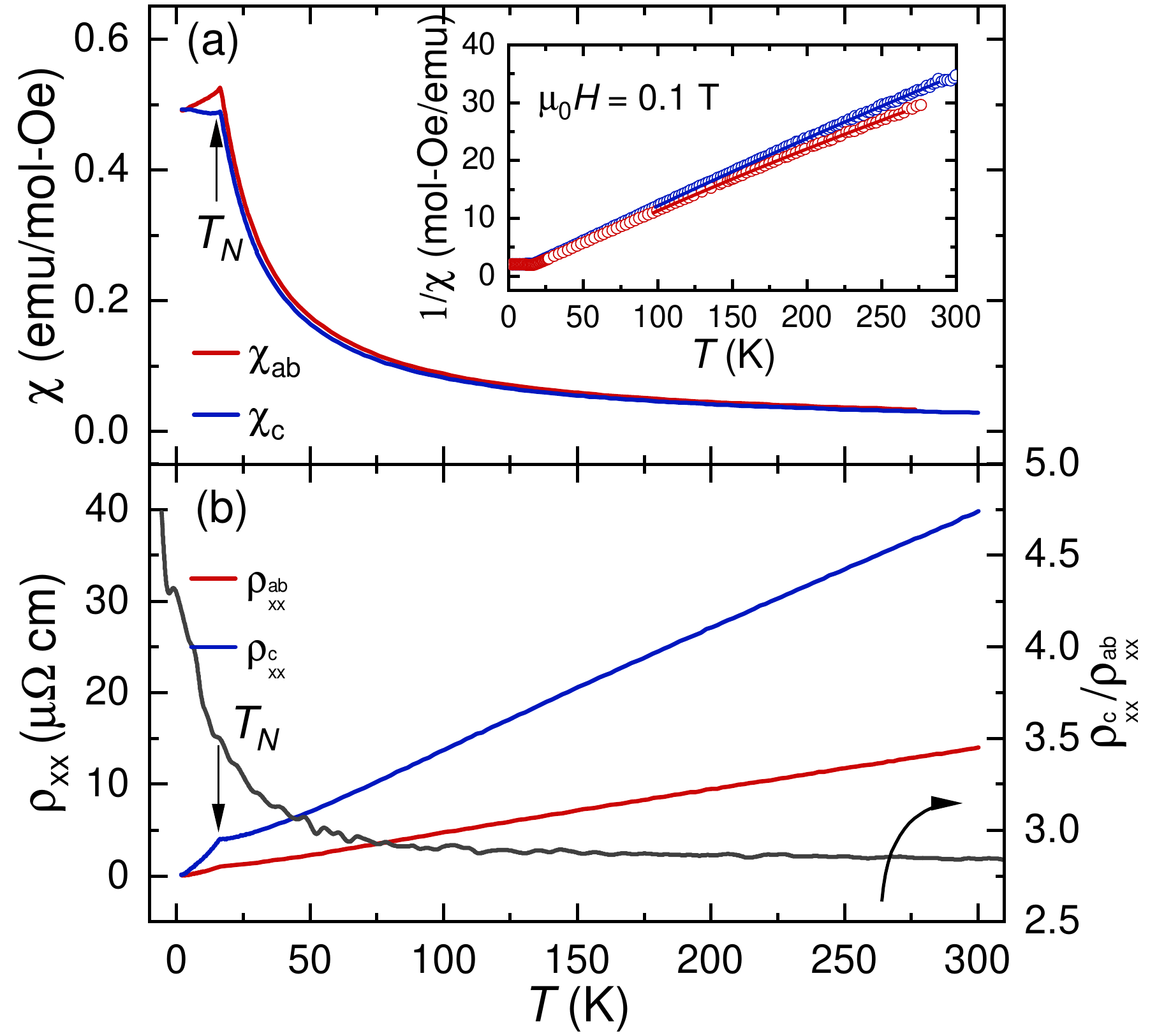}
	\centering
	\vspace{-2ex}%
	\caption{\label{fig:rho_chi}Temperature dependence of the magnetic susceptibility (a) and electrical resistivity (b) for EuGa$_4$. The inset shows the inverse magnetic susceptibility 1/$\chi$ vs. temperature, where the solid lines are fits to the Curie-Weiss model. The arrows indicate an antiferromagnetic transition at $T_\mathrm{N}$ $\sim$ 16.3\,K. The magnetic susceptibility was measured by applying a magnetic field $\mu_0H$ = 0.1\,T, both parallel ($\chi_{c}$) and perpendicular ($\chi_{ab}$) to the $c$-axis. 
	The electrical resistivity was measured in zero field with the current applied both parallel ($\rho_{xx}^{c}$) and perpendicular ($\rho_{xx}^{ab}$) to the  $c$-axis. The ratio of  $\rho_{xx}^{c}$/$\rho_{xx}^{ab}$ is also shown in (b) (right-axis).}
\end{figure*}

Figure~\ref{fig:rho_chi}(a) presents the temperature dependence of the 
dc magnetic susceptibility $\chi(T)$ of EuGa$_4$ between 2 to 300\,K, 
measured in a field of $\mu_0H$ = 0.1\,T, applied both parallel 
($\chi_{c}$) and perpendicular ($\chi_{ab}$) to the 
$c$-axis. For both orientations, as indicated by the arrow in figure~\ref{fig:rho_chi}(a), 
the magnetic susceptibility exhibits a clear anomaly  around  $T_\mathrm{N} \sim 16.4$\,K, which corresponds to the AFM transition of Eu 4$f$ electrons. 
The zero-field-cooling- and field-cooling magnetic susceptibilities are 
practically identical, confirming the AFM nature of the magnetic transition. 
Both $\chi_c$ and $\chi_{ab}$ are comparable in the AFM state and are almost identical in the paramagnetic (PM) state,
indicating the lack of a strong magnetic anisotropy in EuGa$_4$.  
The magnetic susceptibility in the PM state can be described by the modified Curie–Weiss model, $\chi(T) = \chi_0 + C/(T - \theta_p$), with $\chi_0$ being a temperature
independent susceptibility, including contributions from the
core diamagnetism, the van Vleck paramagnetism, and the Pauli paramagnetism; 
$C$ is the Curie constant and $\theta_p$ the PM Curie temperature. 
The inset of figure~\ref{fig:rho_chi}(a) shows the inverse 
susceptibility $1/\chi(T)$ versus $T$ for both $\chi_{c}$ and $\chi_{ab}$, 
respectively, with the solid lines being fits to the  
Curie–Weiss model over the temperature range 150–-300\,K. The slope of 
the fits, allows the determination of the effective magnetic 
moment $\mu_\mathrm{eff}$ from the Curie constant. For $\chi_{c}$, 
the $\mu_\mathrm{eff} = 7.90(5)$\,$\mu_\mathrm{B}$, $\chi_0 =  2.6(2) \times 10^{-3}$\,emu/mol-Oe, 
and $\theta_p = 2(1)$\,K ; while for $\chi_{ab}$, the $\mu_\mathrm{eff} = 7.95(5)$\,$\mu_\mathrm{B}$, 
$\chi_0 = 4.7(3) \times 10^{-3}$\,emu/mol-Oe, and $\theta_p = 5(1)$\,K. 
For both orientations, the effective moments are comparable to the 
theoretical value for free Eu$^{2+}$ ions (7.94~$\mu_\mathrm{B}$).

%
%

The temperature-dependent zero-field electrical resistivity, 
$\rho_{xx}(T)$, measured between 2 and 300 \,K, with the electric current 
applied either parallel ($\rho_{xx}^{c}$) or perpendicular 
($\rho_{xx}^{ab}$) to the $c$-axis, is shown in figure~\ref{fig:rho_chi}(b). 
EuGa$_4$ demonstrates a fairly good metallic behavior. The 
conspicuous difference between $\rho_{xx}^{c}$ and $\rho_{xx}^{ab}$ persists
down to the lowest temperature, with the ratio $\rho_{xx}^{c}$/$\rho_{xx}^{ab}$ reaching $\sim$5 at 2\,K,  
and is consistent with previous 
studies~\cite{nakamura_magnetic_2013}. Such highly anisotropic 
transport properties are most likely attributed to an elongated Fermi 
surface along the $c$-axis~\cite{Kobata2016,nakamura_magnetic_2013} or 
to a different carrier mobility, the latter having been 
proved by Hall measurements in the sister compound EuAl$_4$~\cite{araki_charge_2013}.  
At low temperatures, due to a reduced magnetic scattering, 
both $\rho_{xx}^{c}$ and $\rho_{xx}^{ab}$ exhibit 
a sudden decrease, signaling the onset of an AFM transition of the
Eu 4$f$ electrons.
As indicated by the arrow in figure~\ref{fig:rho_chi}(b), the $T_\mathrm{N}$ determined from electrical resistivity is consistent with that from magnetic susceptibility.

%
\begin{figure}
	\includegraphics[width=0.5\textwidth]{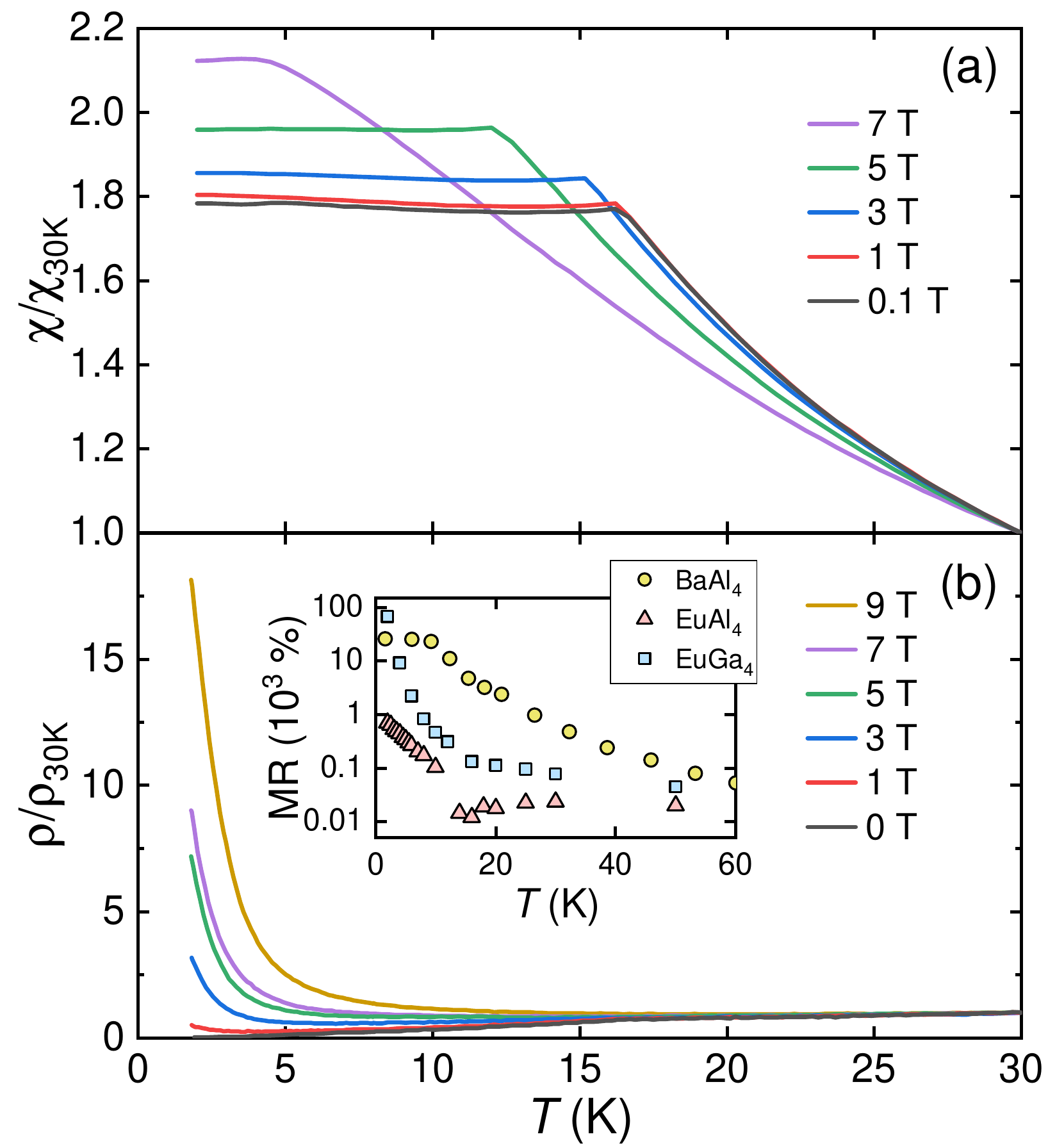}
	\centering
	\vspace{-2ex}%
	\caption{\label{fig:Tdepen}Temperature dependence of the magnetic 
	susceptibility $\chi(T,H)$ (a) and electrical resistivity $\rho_{xx}(T,H)$ (b)  
	for EuGa$_4$, measured in various
	magnetic fields up to 9\,T.  
		Here, the magnetic fields are applied along the $c$-axis. Both 
		the $\chi(T,H)$ and $\rho_{xx}(T,H)$ datasets are normalized 
		to the respective 30-K values. The inset summarizes the 
		magnetoresistance of EuAl$_4$, EuGa$_4$, and of their nonmagnetic 
		 counterpart BaAl$_4$ below 60\,K in a field of 9\,T.
		The magnetoresistance is defined as MR = [$\rho_{xx}(9\mathrm{T}) - \rho_{xx}(0\mathrm{T})$]/$\rho_{xx}(0\mathrm{T})$. Note the logarithmic scale. 
		The data of BaAl$_4$ and EuAl$_4$ were taken from Ref.~\cite{wang_crystalline_2021,EuAl4_PRB}.}
\end{figure}

To further investigate the magnetic and transport properties of EuGa$_4$, both the temperature-dependent magnetic susceptibility $\chi(T,H)$ and electrical resistivity $\rho_{xx}(T,H)$ were measured under various magnetic fields up to 9\,T.
Unlike EuAl$_4$, whose magnetic susceptibility exhibits four successive AFM transitions, only one AFM transition can be identified in the $\chi(T)$ data of EuGa$_4$. 
As can be seen in figure~\ref{fig:Tdepen}(a), upon increasing the 
magnetic field, the AFM transition is progressively suppressed towards 
lower temperatures, reaching $T_\mathrm{N}$ $\sim$ 4.2\,K at 7\,T. 
As for $\rho_{xx}(T,H)$ in figure~\ref{fig:Tdepen}(b), 
the AFM transition can still be identified for applied 
magnetic fields below 5\,T [see also zero-field resistivity 
in figure~\ref{fig:rho_chi}(b)], yet it becomes less visible in  
magnetic fields above 5\,T, because of the onset of a giant MR
at low temperatures. Below 20\,K, the magnetoresistance of 
EuGa$_4$ increases exponentially, reaching almost $\sim 7 \times 10^4$~\% at 2\,K, a record among 
the materials belonging to the BaAl$_4$-type family. 
The MRs vs.\ temperature of EuAl$_4$, EuGa$_4$, and of 
their nonmagnetic counterpart BaAl$_4$ are 
summarized in the inset of figure~\ref{fig:Tdepen}(b). 
Neither EuAl$_4$, nor EuGa$_4$ exhibits 
clear anomalies at the respective AFM transitions.   
Considering that the nonmagnetic BaAl$_4$ also shows a giant 
MR at low temperatures~\cite{wang_crystalline_2021}, we believe 
that the magnetic order of Eu$^{2+}$ ions shows weak correlation to the appearance of such a giant nonsaturating MR in EuAl$_4$ and EuGa$_4$.
Most likely their prominent MR can be attributed to a 
modification of the Fermi surface topology 
by the applied magnetic field, similar to that observed in other topological materials~\cite{Ali2014,Liang2015}.

\begin{figure}
	\includegraphics[width=0.5\textwidth]{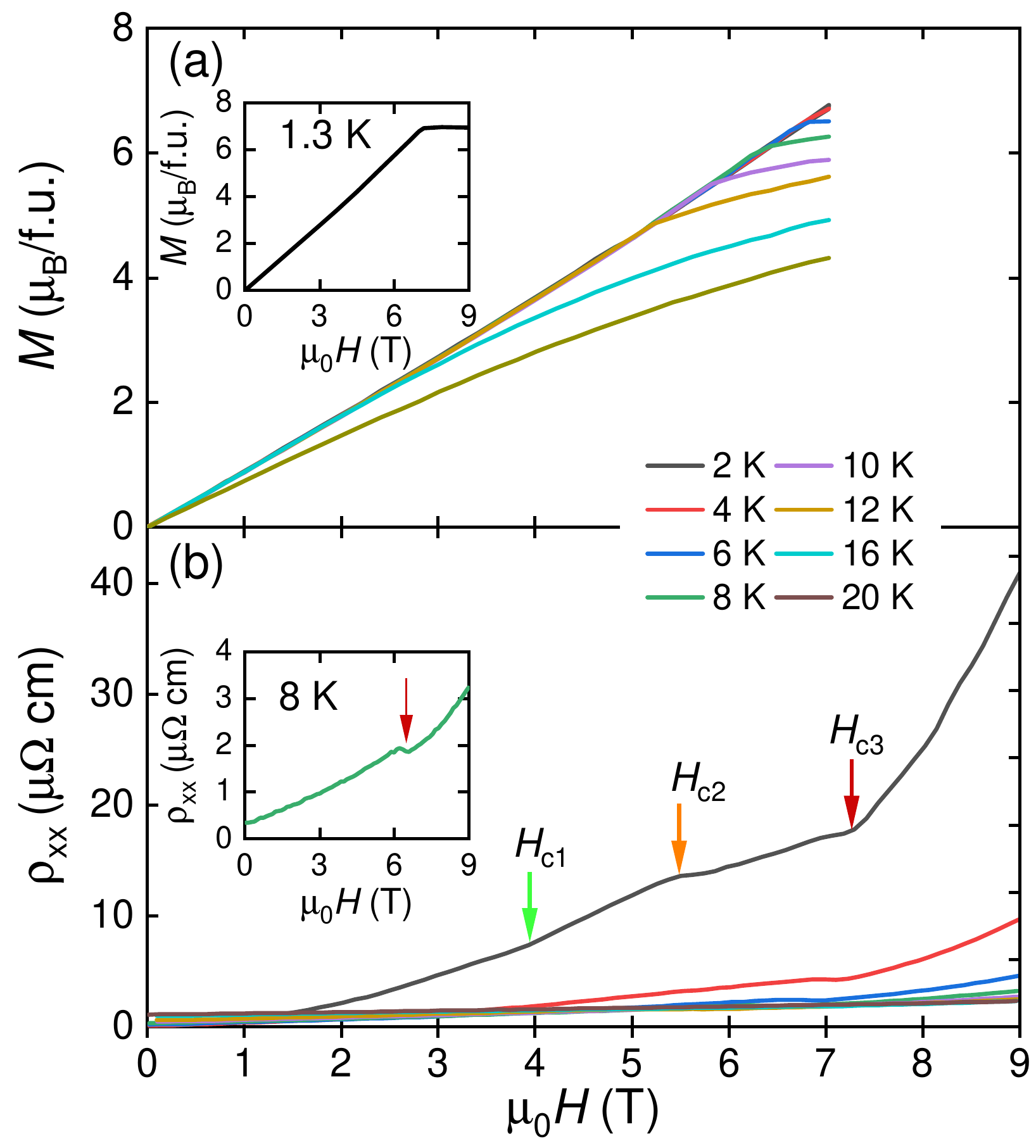}
		\centering
		\vspace{-2ex}%
		\caption{\label{fig:Hdepen}Magnetic field dependence of the 
		magnetization $M(H,T)$ (a) and electrical resistivity $\rho_{xx}(H,T)$ (b) for EuGa$_4$, collected at various temperatures. The magnetic 
		field was applied along the $c$-axis. The arrows in (b) 
		mark the transitions at $H_\mathrm{c1}$, $H_\mathrm{c2}$, 
		and $H_\mathrm{c3}$, respectively. 
		The inset in panel (a) shows $M(H)$ at 1.3\,K, with the data 
		taken from Ref.~\cite{nakamura_magnetic_2013}.
		Inset in (b) shows $\rho_{xx}(H)$ at 8\,K, where only one 
		transition can be identified.}
\end{figure}

Figure~\ref{fig:Hdepen} reports the field dependence of magnetization 
$M(H,T)$ and electrical resistivity $\rho_{xx}(H,T)$ of EuGa$_4$
at various temperatures, covering both the AFM and PM states, 
with the magnetic field applied along the $c$-axis. As shown in 
the inset of figure~\ref{fig:Hdepen}(a), $M(H)$ at 1.3\,K starts to 
saturate for magnetic fields above $\sim$ 7.2\,T, i.e., slightly 
above the highest field we could apply in this study. 
For both field orientations, the saturation magnetization 
$M_s \sim 6.9\,\mu_\mathrm{B}$ is consistent with 7.0\,$\mu_\mathrm{B}$, 
the expected value for the $J = 7/2$ Eu$^{2+}$ ions~\cite{nakamura_magnetic_2013}. 
Such saturation field decreases as the temperature increases,
reaching $\sim 5.3$\,T at 12\,K. 
The sublinear $M(H)$ in the PM states suggests the presence of magnetic 
fluctuations near the AFM order, as have been observed also in EuAl$_4$~\cite{EuAl4_PRB}. 
Unlike the EuAl$_4$ case, whose magnetization shows three 
metamagnetic transitions, accompanied by a small yet clear 
hysteresis in the AFM state, the magnetization of EuGa$_4$ shows 
only a smooth saturation, independent of the direction of the applied 
magnetic field. Surprisingly, when measuring the field-dependent electrical 
resistivity, one can still identify 
three distinct transitions in the AFM state. For example, as shown in figure~\ref{fig:Hdepen}(b), 
$\rho_{xx}(H,2\,\mathrm{K})$ 
undergoes three transitions at $\mu_0H_\mathrm{c1} \sim 3.8$\,T, $\mu_0H_\mathrm{c2} \sim 5.6$\,T, and $\mu_0H_\mathrm{c3} \sim 7.1$\,T,
respectively, which can be clearly tracked also in the first derivative of $\rho_{xx}$
with respect to the magnetic field.
The critical fields $H_\mathrm{c3}$ are consistent with the saturation fields determined from the $M(H)$.
The transitions at $H_\mathrm{c1}$ 
and $H_\mathrm{c2}$ are most likely two metamagnetic transitions, 
similar to those observed also in EuAl$_4$. Why such metamagnetic 
transitions do not show up in the magnetization data requires further investigation. 
Upon increasing the temperature, $H_\mathrm{c2}$ remains 
almost constant and it disappears for $T > 4.5$\,K.
At the same time, $H_\mathrm{c1}$ increases slowly with temperature,
reaching 4.8\,T at 7\,K. Above 7\,K, only one transition 
can be identified in the $\rho_{xx}(H)$ data, as indicated 
by the arrow in the inset of figure~\ref{fig:Hdepen}(b). The $H_\mathrm{c3}$ values determined from electrical-resistivity data
are highly consistent with those from magnetization results, as 
well as with those of previous studies (see magnetic phase diagram below)~\cite{nakamura_magnetic_2013}. 
Note that, both transitions at $H_\mathrm{c1}$ and $H_\mathrm{c2}$
are essentially not reported in previous studies~\cite{nakamura_magnetic_2013}.     

\begin{figure}
	\includegraphics[width=0.5\textwidth]{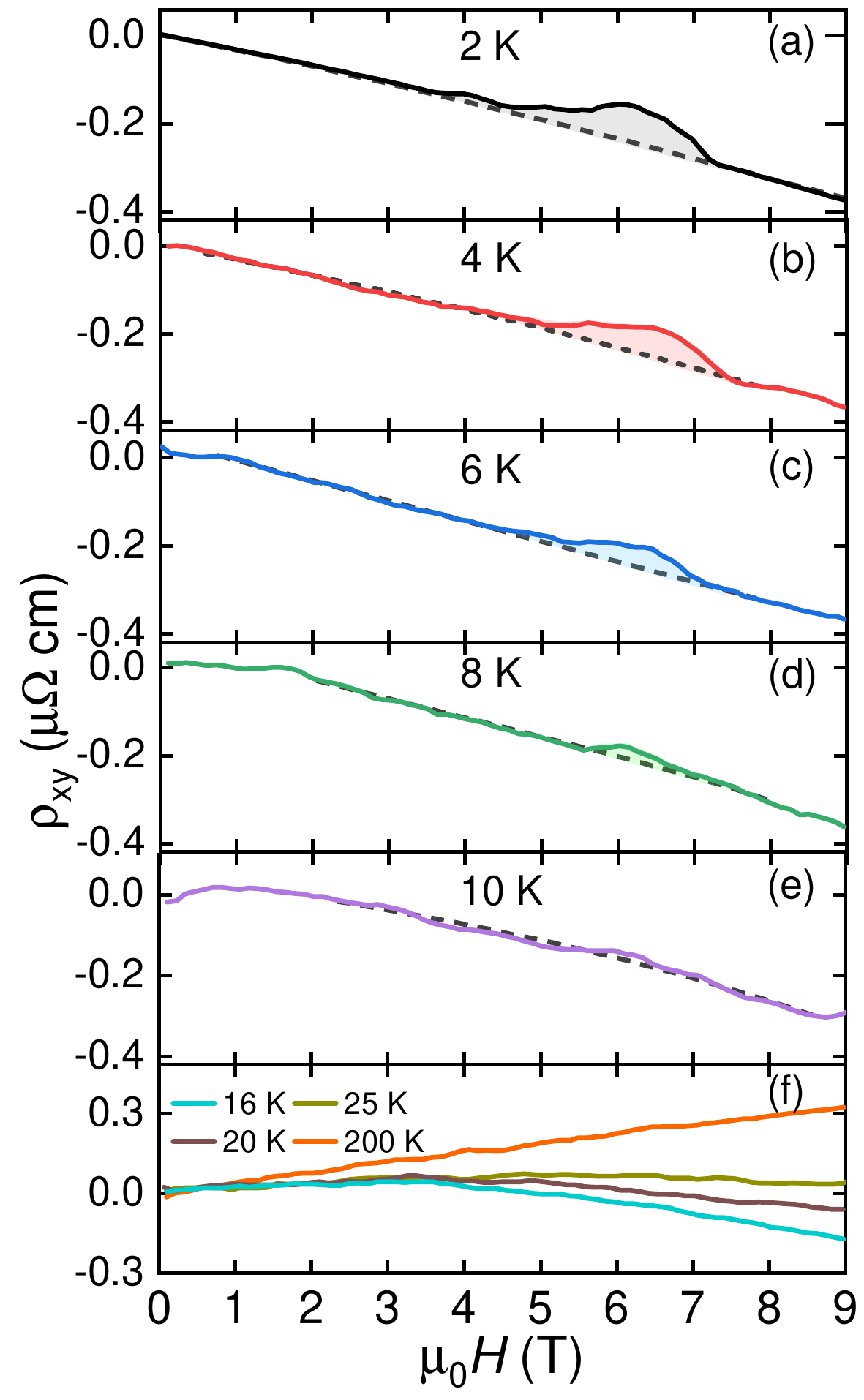}
	\centering
	\vspace{-2ex}%
	\caption{\label{fig:Hall} Magnetic field dependence of the  Hall 
	resistivity $\rho_{xy}(H,T)$ of EuGa$_4$, collected at various 
	temperatures, covering both the AFM and PM states. The magnetic 
	field was applied along the $c$-axis. The dashed lines in (a)-(e) represent polynomial fits.}
\end{figure}

\begin{figure}
	\includegraphics[width=0.58\textwidth]{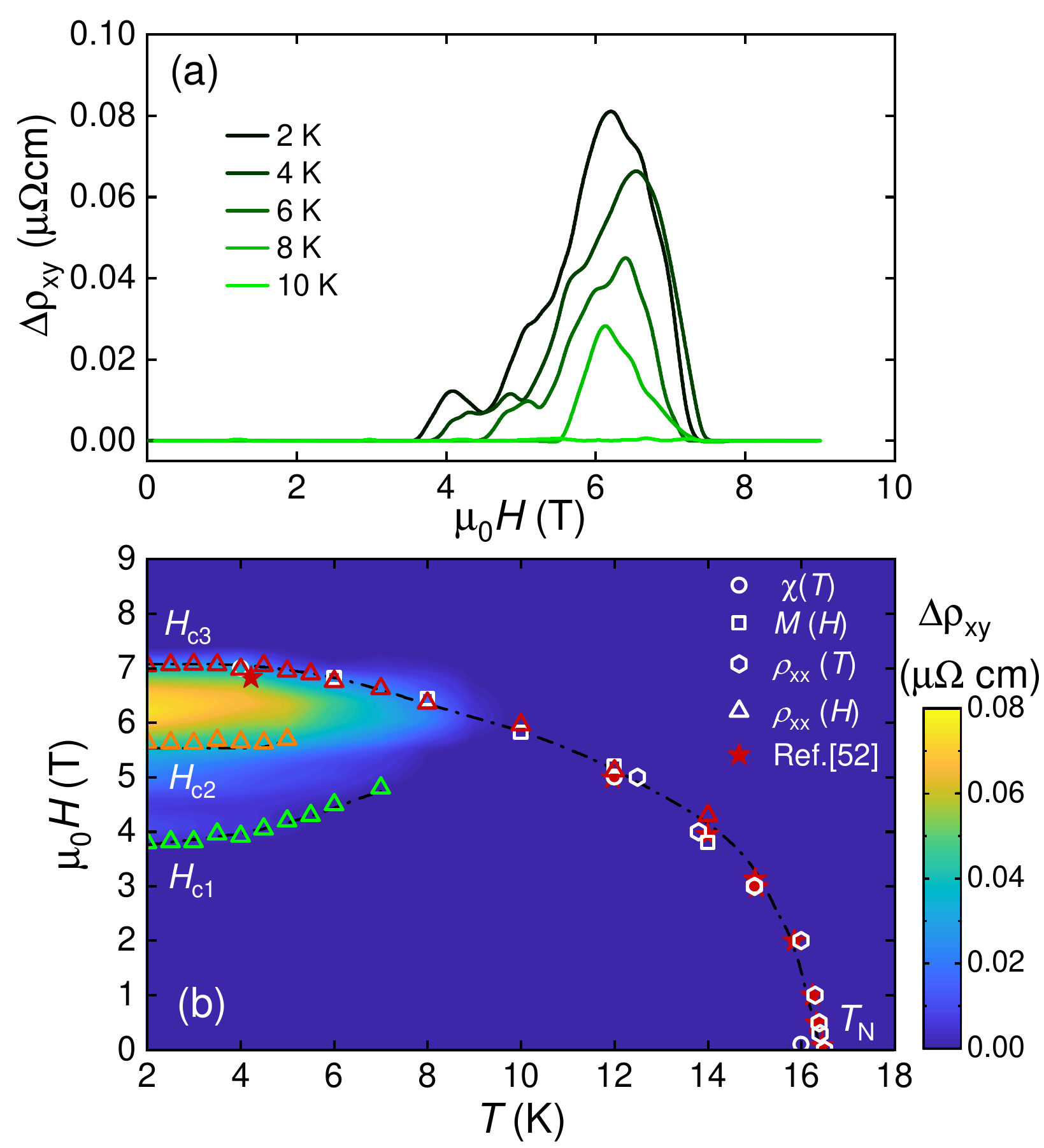}
	\centering
	\vspace{-2ex}%
	\caption{\label{fig:phase}(a) Field dependence of the extracted Hall resistivity $\Delta\rho_{xy}$ of EuGa$_4$ at various temperatures 
		(see text for the definition of $\Delta\rho_{xy}$). 
		(b) Magnetic phase diagram of EuGa$_4$, with the 
		field applied along the $c$-axis. The critical temperatures are determined from $\chi(T)$ and $\rho_{xx}(T)$ under various magnetic fields, while 
		the critical fields are determined from $M(H)$ and 
		$\rho_{xx}(H)$ at various temperatures. The background color in (b) represents 
		the magnitude of $\Delta\rho_{xy}(H)$ at various temperatures. Dash-dotted lines are guides to the eyes. Data taken from Ref.~\cite{nakamura_magnetic_2013} 
		(here shown with stars) are highly consistent with our results.} 
\end{figure}

The field-dependent Hall resistivity $\rho_{xy}(H)$ collected at
various temperatures are presented in figure~\ref{fig:Hall}. In the PM state, as the temperature decreases, the slope of $\rho_{xy}(H)$ changes from positive to negative, implying that EuGa$_4$, too, is a multiband 
system. The multiband nature of EuGa$_4$ is clearly evident from the 
nonlinear behavior of $\rho_{xy}(H)$ [see e.g., $\rho_{xy}(H)$ at 20\,K in figure~\ref{fig:Hall}(f)],  
as confirmed also by de Haas–van Alphen (dHvA) studies 
and by electronic band-structure calculations~\cite{nakamura_unique_2014,nakamura_transport_2015,nakamura_magnetic_2013}.
Similar to BaAl$_4$ and EuAl$_4$~\cite{wang_crystalline_2021,EuAl4_PRB,araki_charge_2013}, 
at high temperatures, the $\rho_{xy}(H)$ of EuGa$_4$ is dominated 
by the holes 
while, at low temperatures, mostly the electrons 
account for the Hall signal.
In the AFM state [see figures~\ref{fig:Hall}(a)-(d)],
a hump-like anomaly in the $\rho_{xy}(H)$ gradually develops with decreasing temperature, reminiscent of the topological Hall resistivity arising from topological spin textures~\cite{neubauer_topological_2009,gayles_dzyaloshinskii-moriya_2015,kanazawa_large_2011,franz_real-space_2014,schulz_emergent_2012,qin_emergence_2019,matsuno_interface-driven_2016,kurumaji_skyrmion_2019,kanazawa_critical_2016,fujishiro_topological_2019,gobel_topological_2020,vistoli_giant_2019}. Such anomaly is clearly evident at low temperatures, but it almost disappears at temperatures above 10\,K [see figure~\ref{fig:Hall}(e)]. 
To determine unambiguously the topological contribution $\rho_{xy}^T$, the ordinary ($\rho_{xy}^O$) and the conventional anomalous ($\rho_{xy}^{A'}$) contributions have to be subtracted from the measured $\rho_{xy}$. 
In the EuGa$_4$ case, its multiband nature makes the subtraction of $\rho_{xy}^O$ unreliable.
Since the hump-like Hall resistivity feature appears only 
in a narrow field range, to extract this anomaly from the 
$\rho_{xy}(H)$ data, we chose to subtract a polynomial 
background [see the dotted lines in figures~\ref{fig:Hall}(a)-(d)]. 
Note that $\Delta\rho_{xy}$ is part of the 
anomalous Hall resistivity, i.e., it might be either trivial (conventional 
anomalous Hall resistivity $\rho_{xy}^{A'}$) or nontrivial (topological 
Hall resistivity $\rho_{xy}^T$).
The subtracted $\Delta\rho_{xy}(H)$ of EuGa$_4$ are plotted in 
figure~\ref{fig:phase}, together with its magnetic phase diagram.
Independent of its nature, clearly $\Delta\rho_{xy}$ is most prominent 
in the AFM state, at temperatures below 10\,K and in a field 
range between $H_\mathrm{c1}$ and $H_\mathrm{c3}$, with 10\,K
representing the critical temperature below which $H_\mathrm{c1}$ 
sets in.  
 
First, we discuss the giant and nonsaturating positive magnetoresistance of EuGa$_4$. Large nonsaturating MR seems 
increasingly ubiquitous in semimetals~\cite{liang_ultrahigh_2015,Ali2014,Wang_three_2013,Huang_observation_2015,tafti_resistivity_2016}. The two most frequently discussed scenarios for explaining the large MR of semimetals are: (i) the charge compensation in a semiclassical two-carrier model and (ii) the guiding center motion of charge carriers. 
The latter is usually relevant for topological semimetals 
with a linear dispersion (see e.g., Ref~\cite{leahy_nonsaturating_2018}). 
In the nonmagnetic BaAl$_4$, the charge compensation is not yet achieved and its large nonsaturating MR is 
associated with the Dirac spectrum~\cite{wang_crystalline_2021}. EuGa$_4$ exhibits comparable MR to the BaAl$_4$, reaching $\sim 7 \times 10^4$\,\% [see details in the inset of figure~\ref{fig:Tdepen}(b)].
Previous dHvA studies and electronic band structure calculations proposed that the charge compensation might be achieved in the PM state of EuGa$_4$~\cite{nakamura_magnetic_2013}.
In the AFM state, the MR increases exponentially, making the estimation of the carrier type and density from the Hall resistivity of a two-band model unreliable.  
Although a possibility of charge compensation cannot be fully excluded, the comparison with BaAl$_4$ suggests a topological nontrivial origin of the giant and nonsaturating MR in EuGa$_4$.
Furthermore, in most of the nonmagnetic semimetals, apart from quantum oscillations, the MR exhibits a linear ($\sim$$H$) or a quadratic ($\sim$$H^2$) magnetic field dependence. 
Obviously, the anomalies at the three critical fields in the MR
of EuGa$_4$ are not related to quantum oscillations [see figure~\ref{fig:Hdepen}(b)].
Considering that a finite $\Delta\rho_{xy}$---possibly signaling topological spin textures---appears mostly in the field range between $H_\mathrm{c1}$ and $H_\mathrm{c3}$ [see figure~\ref{fig:phase}(b)], 
the anomalies at $H_\mathrm{c1}$ and $H_\mathrm{c2}$ are expected to be metamagnetic transitions. At the same time, the critical field $H_\mathrm{c3}$ is consistent with the saturation field in the magnetization data. 
Such metamagnetic transitions might be related to a subtle change of spin direction or magnetic vectors, but their absence in the magnetization data is puzzling and requires further investigation.
A combination of charge compensation, nontrivial band topology, and topological spin textures, may lead to these interesting features in the magnetoresistance of EuGa$_4$.
	
Now we discuss the origin of the hump-like anomaly in $\rho_{xy}(H)$. To check whether a nonzero topological Hall resistivity $\rho_{xy}^T$ underlies the hump-like anomaly in $\rho_{xy}(H)$, a knowledge of the exact field evolution of the ordinary $\rho_{xy}^O(H)$- and conventional anomalous $\rho_{xy}^{A'}(H)$ Hall contributions is crucial. As discussed above, $\rho_{xy}^O(H)$ is unknown a priori, but it is presumably a nonlinear function of field due to the multiband nature of EuGa$_4$.
Furthermore, the extraction of conventional anomalous Hall resistivity $\rho_{xy}^{A'}$ is even more complex in EuGa$_4$. Conventionally, $\rho_{xy}^{A'}$ can be rewritten as $R_sM$, $S_H\rho_{xx}^2M$, or $S_H'\rho_{xx}M$, where $R_s$, $S_H$ and $S_H'$ are constants, and $M$ and $\rho_{xx}$ are the field-dependent magnetization and electrical resistivity, respectively. 
In real materials, $\rho_{xy}^{A'}$ depends on the mechanisms of intrinsic-, side-jump-, skew scattering-, or an intricate combination thereof~\cite{nagaosa_anomalous_2010,tian_proper_2009,hou_multivariable_2015}. 
These different representations together with the multiband nature of EuGa$_4$ make the extraction of $\rho_{xy}^T$ from the measured $\rho_{xy}$ even more complicated, especially considering the presence of a giant MR
in the AFM state of EuGa$_4$. Different methods have been employed to evaluate
$\rho_{xy}^{A'}(H)$ in EuAl$_4$, leading to different explanations for the hump-like anomaly in $\rho_{xy}(H)$. Consequently, the observed $\Delta\rho_{xy}$ in figure~\ref{fig:phase}(a) may correspond exactly to a topological Hall term $\rho_{xy}^T$, or to the lower/upper limits of $\rho_{xy}^T$~\cite{EuAl4_PRB}.
For EuGa$_4$, although a quantitative extraction of $\rho_{xy}^T$ is not yet feasible, it seems that the hump-like anomaly between $H_\mathrm{c1}$ and $H_\mathrm{c3}$ cannot be reproduced  either by a multi-band $\rho_{xy}^O(H)$ (which should evolve smoothly with magnetic field), or by a $\rho_{xy}^{A'}$ in the form of $R_sM$, $S_H\rho_{xx}^2M$, or $S_H'\rho_{xx}M$. Therefore, a nonzero $\rho_{xy}^T$, closely related to the topological spin textures, should be present in EuGa$_4$.
To confirm such topological magnetic phase in EuGa$_4$, further experiments, 
as e.g., resonant x-ray scattering or Lorentz transmission electron microscopy, are highly desirable.
	
The observation of a topological Hall effect in magnetic materials is usually attributed to noncoplanar spin textures, such as magnetic skyrmions, characterized by a finite scalar spin chirality in real space.
Magnetic skyrmions can be stabilized by the Dzyaloshinskii-Moriya interaction and, thus, are often observed in noncentrosymmetric magnetic materials and magnetic thin films~\cite{muhlbauer_skyrmion_2009,yu_near_2011,yu_real-space_2010,seki_observation_2012,kezsmarki_ne-type_2015,tokunaga_new_2015,Seki2012}.
Alternatively, skyrmions can be also stabilized by magnetic frustrations and fluctuations or by the competition between the magnetic interactions and magnetic anisotropies in centrosymmetric systems,
where the Dzyaloshinskii-Moriya interaction
cannot exist~\cite{kurumaji_skyrmion_2019,Hirschberger2019,Khanh2020,Ghimire2020,Batista2016,li_large_2019}. 
Skyrmions in centrosymmetric materials exhibit the unique advantage of
a tunable skyrmion size and spin helicity and are hence intensively pursued~\cite{yu_biskyrmion_2014}.
However, to date, only a few centrosymmetric systems have been found to host skyrmions, e.g., Gd$_2$PdSi$_3$~\cite{kurumaji_skyrmion_2019}, Gd$_3$Ru$_{4}$Al$_{12}$~\cite{Hirschberger2019}, GdRu$_2$Si$_2$~\cite{Khanh2020}, Fe$_3$Sn$_2$~\cite{li_large_2019}, and, possibly, also EuCd$_2$As$_2$ and EuAl$_4$~\cite{xu_unconventional,EuAl4_PRB}.
The observation of a topological Hall effect in EuGa$_4$ not only adds another rare-earth-based skyrmion material, but might also provides clues to the tuning of topological spin textures and other ordered states.
For EuAl$_4$, the THE appears in a field range of 1 to 2\,T~\cite{EuAl4_PRB}.  While in EuGa$_4$, as shown in figure~\ref{fig:phase}(b), the THE shows up between 4 and 7\,T. Therefore, the critical field to induce the THE in EuAl$_4$ is about 3 times smaller than in EuGa$_4$.
On the other hand, a CDW order sets in at
$T_\mathrm{CDW} \sim 140$\,K at ambient pressure in EuAl$_4$~\cite{EuAl4_PRB,araki_charge_2013,nakamura_unique_2014,nakamura_transport_2015,shimomura_lattice_2019,Kobata2016}, whereas physical- or chemical pressure is necessary to induce the CDW order in EuGa$_4$~\cite{nakamura_transport_2015,nakamura_magnetic_2013,Stavinoha2018}. 
The $T_\mathrm{CDW}$ decreases with increasing pressure in EuAl$_4$,
while it does exactly the opposite in EuGa$_4$~\cite{nakamura_transport_2015,nakamura_magnetic_2013}.
At low temperatures, in both EuAl$_4$ and EuGa$_4$, $T_\mathrm{N}$
increases with increasing pressure, exceeding 30\,K at a pressure of 5\,GPa in EuAl$_4$~\cite{nakamura_transport_2015,nakamura_magnetic_2013}. 
Future studies on the evolution of $\rho_{xy}^T$ with pressure in Eu(Ga$_{1-x}$Al$_x$)$_4$ could shed light on the exotic magnetic phase with topological spin
textures of such systems.
	
In addition to the topological spin textures, upon breaking certain symmetries, noncollinear antiferromagnets may also exhibit a topological Hall effect due to Berry curvature, such as at the Dirac or Weyl points~\cite{THEnote}. Such a momentum-space scenario has been theoretically proposed and experimentally observed, for instance, in Mn$_3$(Sn,Ge)~\cite{nakatsuji_large_2015,nayak_large_2016,ikhlas_large_2017} and (Yb,Gd)PtBi~\cite{suzuki_large_2016,guo_evidence_2018}.
A three-dimensional Dirac spectrum with nontrivial topology and possible nodal-lines crossing the Brillouin zone has been recently observed in nonmagnetic BaAl$_4$~\cite{wang_crystalline_2021}. 
The giant nonsaturating MR in EuGa$_4$ hints at the
existence of a topologically nontrivial band structure,
which may cooperate with topological spin textures and, thus, 
contribute to the topological Hall effect.

\section{Summary} 
To summarize, we observed a giant nonsaturating magnetoresistance and a hump-like anomaly $\Delta\rho_{xy}$ in the Hall resistivity of the centrosymmetric antiferromagnet EuGa$_4$. 
Analogous to the nonmagnetic BaAl$_4$, the MR of EuGa$_4$ could originate from its nontrivial band topology. 
By systematic temperature- and field-dependent electrical resistivity and magnetization measurements, we could establish the magnetic phase diagram of EuGa$_4$. Similarly to EuAl$_4$, the hump-like anomaly in the Hall resistivity 
of EuGa$_4$ appears mostly in a field range where also metamagnetic transitions occur.  
Such hump-like anomaly is most likely an indication of the topological Hall effect.
Although a trivial origin of the effect cannot be fully excluded, our results suggest that a topological Hall effect and topological spin textures, such as magnetic skyrmions, may indeed exist in EuGa$_4$. 
Therefore, the material family with a tetragonal BaAl$_4$-type structure, 
to which EuGa$_4$ and EuAl$_{4}$ belong, seems to comprise suitable candidates
on which one can study the interplay between correlated-electron phenomena
(such as charge-density wave or exotic magnetism) with topological spin
textures and topologically nontrivial band structures.

\section{Acknowledgements}	
T.S.\ acknowledges support from the Natural Science Foundation of 
Shanghai (Grant Nos.\ 21ZR1420500 and 21JC1402300) and the Schweizerische Nationalfonds 
zur F\"{o}rderung der Wis\-sen\-schaft\-lichen For\-schung (SNF) (Grant Nos.\ 200021\_188706 
and 206021\_139082). Y.X.\ acknowledges support from the Shanghai Pujiang Program (Grant No.\ 21PJ1403100).
This work was also financially supported by the National Natural Science foundation of China (NSFC) (Grant Nos. 12174103 and 11874150) 
and the Sino-Swiss Science and Technology Cooperation (Grant No.\ IZLCZ2-170075).
\section{References}
	
\bibliography{EuGa4}	
	
\end{document}